\documentstyle[12pt]{article}

\def\d{^\dagger}
\def\bra#1{\langle #1 |}
\def\ket#1{| #1 \rangle}
\def\expect#1{\langle #1 \rangle}
\def\Tr{{\rm Tr}}

\begin{document}

\title{A C++ library using quantum trajectories \\
to solve quantum master equations}

\author{R\"udiger
  Schack\thanks{Email: r.schack@rhbnc.ac.uk} $^{^{\hbox{\tiny(a,b)}}}$ 
and Todd A. Brun\thanks{Email: t.brun@qmw.ac.uk} $^{^{\hbox{\tiny(b)}}}$   
    \vspace{3mm}\\
$^{\hbox{\tiny(a)}}${\it Department of Mathematics, 
             Royal Holloway, University of London} \\
             {\it Egham, Surrey TW20 0EX, England}  \vspace{3mm}\\
$^{\hbox{\tiny(b)}}${\it Department of Physics, 
             Queen Mary and Westfield College} \\
             {\it University of London, London  E1 4NS, England}}

\date{July 18, 1996}
\maketitle

\begin{abstract}
Quantum trajectory methods can be used for a wide range of open quantum systems
to solve the master equation by unraveling the density operator evolution into
individual stochastic trajectories in Hilbert space.  This C++ class library
offers a choice of integration algorithms for three important unravelings of
the master equation. Different physical systems are modeled by different
Hamiltonians and environment operators. The program achieves flexibility and
user friendliness, without sacrificing execution speed, through the way it
represents operators and states in Hilbert space. Primary operators,
implemented in the form of simple routines acting on single degrees of freedom,
can be used to build up arbitrarily complex operators in product Hilbert spaces
with arbitrary numbers of components. Standard algebraic notation is used to
build operators and to perform arithmetic operations on operators and states.
States can be represented in a local moving basis, often leading to dramatic
savings of computing resources. The state and operator classes are very general
and can be used independently of the quantum trajectory algorithms. Only a
rudimentary knowledge of C++ is required to use this package.
\end{abstract}

%\vspace{3mm}

%\begin{tabbing}
%{\bf Corresponding author:} \= Dr.~R. Schack \\
% \> Department of Mathematics \\
% \> Royal Holloway, University of London \\
% \> Egham, Surrey TW20 0EX \\
% \>{\bf Tel.} 01784 443097 \\
% \>{\bf FAX} 01784 430766 \\
% \>{\bf Email} r.schack@rhbnc.ac.uk
%\end{tabbing}

\newpage

\section*{Program Summary}

{\it Title of program\/}: Quantum trajectory class library 

\vspace{3mm}\noindent
{\it Program obtainable from\/}: 
http://galisteo.ma.rhbnc.ac.uk/applied/QSD.html and the authors.

\vspace{3mm}\noindent
{\it Licensing provisions\/}: none

\vspace{3mm}\noindent
{\it Operating systems under which the program has been tested\/}:
UNIX (Gnu g++), DOS (Turbo C++), VMS (DEC C++)

\vspace{3mm}\noindent
{\it Programming language used\/}: C++

\vspace{3mm}\noindent
{\it Memory required to execute with typical data\/}: 1MByte

\vspace{3mm}\noindent
{\it Has the code been vectorized?\/}: no

\vspace{3mm}\noindent
{\it No.~of lines in distributed program, including test data, etc.}: 8000

\vspace{3mm}\noindent {\it Keywords\/}: open quantum system, master equation,
Hilbert space, quantum trajectories, unraveling, stochastic simulation, quantum
computation, quantum optics, quantum state diffusion, quantum jumps, Monte
Carlo wavefunction

\vspace{3mm}\noindent
{\it Nature of physical problem\/}:\\
Open quantum systems, i.e., systems whose interaction with the environment
can not be neglected, occur in a variety of contexts. Examples are quantum
optics, atomic and molecular physics, and quantum computers. If the time
evolution of the system is approximately Markovian, it can be described by a
master equation of Lindblad form [1], a first order differential equation for
the density operator. Solving the master equation is the principal purpose of
the program. Since the state and operator classes are very general, they can
be used in any physical problem involving Hilbert spaces with several degrees
of freedom.

\vspace{3mm}\noindent
{\it Method of solution\/}:\\
By analogy with the solution of a Fokker-Planck equation by numerical 
simulation of the corresponding stochastic differential equation, a master
equation can be solved by simulating the stochastic evolution of a vector in
Hilbert space. The correspondence between master
equation and stochastic equation is not unique: there are many ways to {\it
unravel\/} the master equation into stochastic quantum trajectories. The program
implements three such unravelings, known as the ``quantum state diffusion
method (QSD)'' [2],  the ``quantum jump method'' [3--5], and the ``orthogonal
jump method''[6]. The phenomenon of
phase-space localization [7,8] is exploited numerically by representing quantum
states in a local moving basis obtained by applying the coherent-state
displacement operator to the usual harmonic-oscillator basis, often leading to 
dramatic savings of computing resources.

\vspace{3mm}\noindent {\it Unusual features of the program\/}:\\ It is worth
emphasizing the effortless way in which operators and states in product Hilbert
spaces are represented. Primary operators implemented in the form of simple
routines acting on single degrees of freedom can be used to build up
arbitrarily complex operators in product Hilbert spaces with arbitrary numbers
of components. Building operators, performing arithmetic operations on
operators and states, and applying operators to states is done using standard
algebraic notation. This program structure has been made possible by
systematically implementing object-oriented programming concepts such as
inheritance, concepts which are not (yet) widely used in computational
physics. Encapsulation of program modules makes it easy to add new basic
operators, alternative unravelings of the master equation, or different
integration algorithms.

\vspace{3mm}\noindent
{\it Typical running time\/}:\\
The running time depends on the complexity of the problem, the integration time,
and the number of trajectories required. A typical running time for a simple
problem is a few minutes. There is no upper limit.

\vspace{3mm}\noindent
{\it References\/}:

\noindent[1] G. Lindblad, Commun.\ Math.\ Phys.\ {\bf 48},  119  (1976).

\noindent[2] N. Gisin and I.~C. Percival, J. Phys.\ A {\bf 25},  5677  (1992).

\noindent[3] H.~J. Carmichael, {\em An Open Systems Approach to Quantum
  Optics} (Springer, Berlin, 1993).

\noindent[4] J. Dalibard, Y. Castin, and K. M{\o}lmer, Phys.\ Rev.\ Lett.\
  {\bf 68}, 580 (1992).

\noindent[5] C.~W. Gardiner, A.~S. Parkins, and P. Zoller, Phys.\ Rev.\ A {\bf 46},  4363
  (1992).

\noindent[6] L. Di\'osi, Phys.\ Lett.\ A {\bf 114},  451  (1986).

\noindent[7] T. Steimle, G. Alber, and I.~C. Percival, 
  J. Phys.\ A {\bf 28}, L491 (1995).

\noindent[8] R. Schack, T. A. Brun, and I. C. Percival,
  J. Phys.\ A {\bf 28}, 5401 (1995).

\section*{Long Write-Up}

\section{Introduction}

For many quantum systems of current interest it is no longer possible
to neglect the interactions with the environment. Those
so-called {\it open quantum systems\/} occur in a variety of contexts
including quantum optics, atomic and molecular physics, and quantum
computers. Open quantum systems can often be described by a master
equation {\cite{Lindblad1976}}, a first-order differential equation
for the density operator, in which the internal dynamics of the 
system is represented by the system Hamiltonian $\hat H$, which is a
Hermitian Hilbert-space operator, and the interaction with the
environment is represented by one or more Lindblad operators $\hat L_j$
which are not necessarily Hermitian.

By analogy with the solution of a Fokker-Planck equation by numerical simulation
of the corresponding stochastic differential equation (or Langevin equation), a
master equation can be solved by simulating the stochastic evolution of a
vector in Hilbert space. The correspondence between
master equation and stochastic equation is not unique; there are many ways to
{\it unravel\/} the master equation into stochastic quantum trajectories
{\cite{Diosi1986,Gisin1992c,%
Carmichael1993b,Dalibard1992,Gardiner1992,Breslin1995}}.

The main challenge of this software project was to develop a general
program flexible enough to accommodate different integration
algorithms and unravelings of the master equation, as well as the vast
range of possible physical systems. In particular, we wanted to make
it easy to add new algorithms and unravelings, and we wanted a
program capable of dealing with arbitrary Hamiltonian and
Lindblad operators in Hilbert spaces with an arbitrary number of degrees
of freedom. This task turned out to be ideal for the
application of object-oriented programming. We chose the C++
language both because of its wide availability and because it allowed us to
use standard mathematical notation for Hilbert-space operations by
overloading algebraic operators like `+' and `$\ast$'.

The core of the program are the C++ classes {\tt State} and
{\tt Operator}, which represent state vectors and operators in Hilbert space.
Because of the object-oriented features of C++, it is possible to hide the
implementation details of these classes completely from the classes dealing
with the simulation of quantum trajectories. These implementation details
need not to be known either by a user of the program who wants to choose the
quantum operators defining the physical problem of interest or by a
programmer who wants to add a new unraveling of the master equation to the
software. A welcome side effect of this encapsulation is that the
{\tt State} and {\tt Operator} classes can be used independently
of the rest of the code. They
should prove useful in many numerical schemes involving Hilbert spaces  for
systems with several degrees of freedom.

Many Hamiltonian and Lindblad operators can be written as
sums of products of simple operators acting on a single degree of
freedom. Here is an example of a Hamiltonian operator coupling a two-level
atom (with raising and lowering operators $\hat\sigma_+$ and $\hat\sigma_-$) to
an electromagnetic field mode (with annihilation and creation operators 
$\hat a$ and $\hat a\d$):
\begin{equation}
\hat H = g\left(\hat\sigma_+ \hat a + \hat\sigma_- \hat a\d \right) \;,
\end{equation}
where the parameter $g$ is the coupling strength.
In the following code segment, the atomic and field degrees of freedom are
labeled 0 and 1, respectively. The Hamiltonian is defined in terms of the
predefined {\it primary operators\/} {\tt SigmaPlus} and
{\tt AnnihilationOperator} using standard algebraic notation. The class
{\tt AdaptiveStep} is a {\it stepper\/} routine advancing the quantum trajectory
by a single time step.
\begin{verbatim}
double g = 0.5;
SigmaPlus Sp(0);             // operates on the 1st degree of freedom
AnnihilationOperator A(1);   // operates on the 2nd degree of freedom
Operator Sm = Sp.hc();       // Hermitian conjugate
Operator Ac = A.hc();
Operator H = g*( Sp*A + Sm*Ac );    // Hamiltonian
 ...
AdaptiveStep theStepper(..., H, ...);  // ... denotes further arguments
\end{verbatim}
The important feature illustrated by this example is that the
stepper routine is passed an object of type {\tt Operator} without any
reference to details like the number of degrees of freedom. 
All the stepper needs to know is that operators can be added, multiplied,
etc., and that they can be applied to state vectors.

Internally, the primary operators {\tt SigmaPlus} and
{\tt AnnihilationOperator} are represented as simple loops acting on a
single-degree-of-freedom state vector. An instance of the more general
{\tt Operator} class is represented by a stack that indicates which
primary operators are used and the operations by which they are combined.
For example, the sequence of steps executed by the program when the operator
$\hat H$ defined above is applied to a state $\ket\psi$ is summarized in the 
expression
\begin{equation}
\hat H \ket\psi = g\left(\hat\sigma_+(\hat a\ket\psi)
        + \hat\sigma_-(\hat a\d\ket\psi) \right) \;,
\end{equation}
in which the  elementary steps are applying a primary operator to a state,
adding two states, and multiplying a state by a scalar. It is clear from this
example that a different grouping of the terms in the expression for $\hat H$
could lead to inefficient code. This will be discussed in Sec.~\ref{secop}.

\section{Quantum trajectories}

\subsection{Master equations}

An open quantum system cannot be described by a
Hilbert-space vector $\ket{\psi}$ evolving according to the
Schr\"odinger equation; instead, the state
must be described by a density operator $\hat\rho$ whose time
evolution generally does not follow any simple law. Fortunately it
turns out that for a large class of systems the time evolution of the
density operator $\hat\rho$ is {\it Markovian\/} to an excellent
approximation, i.e., the rate of change of $\hat\rho$ at time $t$,
$d\hat\rho/dt$, depends only on $\hat\rho(t)$, not on the value of
$\hat\rho$ at any earlier time. It has been shown that under the
Markov approximation the density operator of any open quantum system
obeys 
a {\it master equation\/} of Lindblad form \cite{Lindblad1976}
\begin{equation}
{d\over{dt}}\,\hat\rho =
 -{i\over\hbar}[\hat H,\hat\rho] +
 \sum_j\left(\hat L_j\hat\rho \hat L_j^\dagger
 - {1\over2} \hat L_j^\dagger\hat L_j\hat\rho   - {1\over2}\hat\rho
 \hat L_j^\dagger\hat L_j\right) \;,
\label{eqmaster}
\end{equation}
where $\hat H$ is the system Hamiltonian and the $\hat L_j$ are the
Lindblad operators representing the interaction with the environment.

In many cases, no analytical methods for the solution of the master
equation are known; one has to use numerical methods. But even a
numerical solution of the master equation can be very hard. 
If a state requires $D$ basis vectors
in Hilbert space to represent it, the corresponding density operator
will require $D^2 - 1$ real numbers; this can often be too large for
even the most powerful machines to handle, particularly if the system
involves more than one degree of freedom. 

This problem can be overcome by unraveling the density operator evolution into
{\it quantum trajectories\/} {\cite{Diosi1986,%
Gisin1992c,Carmichael1993b,Dalibard1992,Gardiner1992,Breslin1995}}.  Since
quantum trajectories represent the system as a state vector rather than a
density operator, they often have a numerical advantage over solving the master
equation directly, even though one has to average over many quantum
trajectories to recover the solution of the master equation. A single quantum
trajectory can give an excellent, albeit qualitative, picture of a single
experimental run.

\subsection{Unravelings}    

The three unravelings of the master equation currently implemented are given
by the following three nonlinear stochastic differential
equation for a normalized state vector $|\psi\rangle$: 

\noindent(i) the quantum state diffusion (QSD) equation {\cite{Gisin1992c}}
\begin{eqnarray}
|d\psi\rangle &=& -{i\over\hbar} \hat H \,|\psi\rangle dt 
  + \sum_j\left(\langle \hat L_j^\dagger\rangle_\psi \hat L_j
  - {1\over2} \hat L_j^\dagger \hat L_j 
  - {1\over2} \langle \hat L_j^\dagger\rangle_\psi
  \langle\hat L_j\rangle_\psi\right) |\psi\rangle dt \nonumber \\
&&+\sum_j \left(\hat L_j -\langle \hat L_j\rangle_\psi\right) 
          \,|\psi\rangle d\xi_j \;,
\label{eqqsd}
\end{eqnarray}
(ii) the quantum jump equation \cite{Carmichael1993b,Dalibard1992,Gardiner1992}
\begin{eqnarray}
|d\psi\rangle &=& -{i\over\hbar} \hat H \,|\psi\rangle dt 
+ \sum_j \left(
    {1\over2} \langle \hat L_j^\dagger\hat L_j\rangle_\psi
  - {1\over2} \hat L_j^\dagger \hat L_j \right) |\psi\rangle dt \nonumber\\
&&+ \sum_j \left(
\frac{\hat L_j}{\sqrt{\langle \hat L_j^\dagger\hat L_j\rangle_\psi}}-1
\right)           \,|\psi\rangle dN_j \;,
\label{eqqj}
\end{eqnarray}
and (iii) the orthogonal jump equation \cite{Diosi1986,Breslin1995}
\begin{eqnarray}
|d\psi\rangle &=& -{i\over\hbar} \hat H \,|\psi\rangle dt 
  + \sum_j\left(\langle \hat L_j^\dagger\rangle_\psi \hat L_j
  - {1\over2} \hat L_j^\dagger \hat L_j 
  + {1\over2} \langle \hat L_j^\dagger\hat L_j\rangle_\psi 
  - \langle \hat L_j^\dagger\rangle_\psi
  \langle\hat L_j\rangle_\psi\right) |\psi\rangle dt \nonumber \\
&&+ \sum_j \left(
\frac{\hat L_j - \expect{\hat L_j}_\psi}
     {\sqrt{\expect{\hat L_j\d\hat L_j}_\psi -
                 \expect{\hat L_j\d}_\psi \expect{\hat L_j}_\psi}}
 -1 \right)    \,|\psi\rangle dN_j  \;.
\label{eqortho}
\end{eqnarray}
The first sum in each of these equations represents the deterministic drift of
the state vector due to the environment, and the second sum the random
fluctuations. Angular brackets denote the quantum expectation $\langle\hat
G\rangle_\psi = \langle \psi|\hat G|\psi\rangle$ of the operator $\hat G$ in
the state $|\psi\rangle$. The $d\xi_j$ are independent complex differential
Gaussian random variables satisfying the conditions
\begin{equation}
{\rm M} d\xi_j = {\rm M} d\xi_i d\xi_j = 0\;,\;\;\; 
{\rm M} d\xi_i^* d\xi_j = \delta_{ij}dt \;,
\end{equation}
where {\rm M} denotes the ensemble mean.
The $dN_j$ are independent real discrete Poissonian random variables 
satisfying the conditions
\begin{equation}
dN_j^2=dN_j \;,\;\;\; 
dN_i dN_j = 0\;,\;\;\; 
{\rm M}_{|\psi\rangle} dN_j = 
\left( \langle \hat L_j^\dagger\hat L_j\rangle_\psi 
- \lambda\expect{\hat L_j\d}_\psi \expect{\hat L_j}_\psi\right) \, dt \;,
\end{equation}
where the ``conditional mean'' ${\rm M}_{|\psi\rangle}$ is defined as the mean
over all trajectories for which $|\psi(t)\rangle=|\psi\rangle$, and where
$\lambda=0$ for the quantum jump equation~(\ref{eqqj}) and $\lambda=1$ for the
orthogonal jump equation~(\ref{eqortho}).

The density operator is given by the
mean over the projectors onto the quantum states of the ensemble:
\begin{equation}
\hat\rho = {\rm M}|\psi\rangle\langle\psi| \;.
\end{equation}
If the pure states of the ensemble satisfy one of the quantum trajectory
equations (\ref{eqqsd}), (\ref{eqqj}), or (\ref{eqortho}), then
the density operator satisfies the master equation (\ref{eqmaster}):
\begin{equation}
{\rm M}\ket{\psi(t)} \bra{\psi(t)} = \hat\rho(t),
\end{equation}
where we have assumed that initially the system is in a pure state
$\ket{\psi_0}$ at time $t=0$.  From this it is clear that the expectation
value of an operator $\hat O$ is given by
\begin{equation}
\Tr\{\hat O\hat\rho \} = {\rm M} \bra\psi\hat O \ket\psi  \;.
\end{equation}

\section{Program Structure}

Our C++ library can be divided roughly into three large parts:

1.  The {\tt State} class and its associated friend functions.  A {\tt State}
includes as member data the number of degrees of freedom it represents, how
many basis vectors are allocated for each degree of freedom, the physical type
of each degree of freedom, and (of course) the complex amplitudes of each basis
vector in the total Hilbert space.  The member functions include constructors
for a number of common {\tt State} types; arithmetic functions enabling {\tt
State}s to be added, subtracted, multiplied by scalars, and normalized;
functions relating to the efficient use of memory, so that a {\tt State} can be
dynamically resized; and functions controlling the action of {\tt Operator}s on
the {\tt State}.  There are also member data and functions relating to the
moving basis algorithm, described below. {\tt State}s (and {\tt Operator}s) can
be used like ordinary variables. In particular, when a locally defined {\tt
State} (or {\tt Operator}) goes out of scope, all memory used by it is properly
returned to the system; the user of the program need not worry about memory
allocation and deallocation as this is done automatically.

2.  The {\tt Operator} class.  {\tt Operator}s are defined in terms of their
actions on {\tt State}s.  There is a small class of {\tt PrimaryOperator}s,
whose actions on a single degree of freedom are given by pre-defined
functions.  More complex {\tt Operator}s are defined in terms of these {\tt
PrimaryOperator}s; they can be added, multiplied, multiplied by scalars or
time-dependent functions, conjugated, or raised to powers.  An {\tt Operator}'s
member data includes a number of dynamically allocated stacks which indicate
which {\tt PrimaryOperator}s are used, and the operations by which they are
combined.  Arithmetic operations on {\tt Operator}s are then defined by
operations on these stacks.

3.  The {\tt Trajectory} class and associated classes.  These encode the
numerical algorithms for solving the quantum trajectory equations and
generating output, with associated integration routines, random number
generators, and other utilities.  Several different integration algorithms are
currently included, including second- and fourth-order Runge-Kutta and
Cash-Karp Runge-Kutta with adaptive time steps {\cite{Press1992}}.  These
algorithms are used to solve the deterministic part of the quantum trajectory
equations (\ref{eqqsd}), (\ref{eqqj}), and (\ref{eqortho}).  The stochastic
terms are solved using first-order Euler integration.  The implementation of
more sophisticated stochastic integration methods (see, e.g.,
\cite{Steinbach1995a}) is straightforward. Note that it is only in this part of
the program that there is any reference at all to the details of quantum
unravelings. The {\tt Operator} and {\tt State} classes are very general.

These three parts are roughly equal in size, but quite different in internal
structure.  The {\tt State} class is a single monolithic C++ class with
associated functions; the {\tt Operator} class is a parent class with numerous
descendent classes representing the different {\tt PrimaryOperator}s.  The
numerical integration classes are independent of the details of {\tt
State} and {\tt Operator}, and of each other.  Because of the object-oriented
nature of C++, these three groups need know very little about each other's
internal workings. The following more detailed discussion is not exhaustive; a
complete description of the code can be found in the extensively commented
{\tt \#include} files, particularly in {\tt State.h, Operator.h, and Traject.h}.

\section{The State and Operator Classes}

\subsection{One degree of freedom}

\subsubsection{States}

We represent a state $\ket\psi$ with a single degree of freedom by an
array of $N$ complex amplitudes $c_j$ in a given basis
$\{\ket{\phi_j}\}$:
\begin{equation}
\ket\psi = \sum_{j=1}^N c_j\ket{\phi_j} \;.
\end{equation}
The choice of basis vectors depends on the physical type of the system. For
field modes, we use Fock states $\ket n$; for spins ($s=1/2$), we use
$\hat\sigma_z$ eigenstates $|\!\downarrow\rangle$ and $|\!\uparrow\rangle$; for
$N$-level atoms, we use energy levels $\ket j$. Other types, e.g., molecules
or higher spins, can be added easily. Of course, a true field mode
has an infinite-dimensional Hilbert space.  The {\tt State} class represents
fields by a finite number of basis states, which should be taken as a
truncation of the true infinite expansion.

To represent a state then requires the physical type (currently {\tt FIELD},
{\tt SPIN} or {\tt ATOM}), the number of basis vectors $N$, and an array of $N$
complex amplitudes. The state class contains constructors for many typical
situations. For instance, the expression
\begin{verbatim}
State psi(2,SPIN);
\end{verbatim}
defines {\tt psi} to be the $|\!\downarrow\rangle$ state of a spin ($N=2$), and
\begin{verbatim}
Complex alpha(0.2,0.3); State psi(100,alpha,FIELD);
\end{verbatim}
defines a coherent state $\ket\alpha$ with $\alpha=0.2+0.3i$ truncated to
$N=100$ basis states.

Arithmetic operations for {\tt State}s are defined internally as operations
on the complex amplitudes. In the following code examples, the state
$\ket{\psi_3}=0.5\ket{\psi_1}-\ket{\psi_2}$ is formed from the Fock states
$\ket{\psi_1}=\ket0$ and $\ket{\psi_2}=\ket3$, added to $\ket{\psi_1}$, and
then renormalized; finally, the inner product $z=\langle\psi_2\ket{\psi_3}$ is
evaluated. Here $N=10$ basis states are more than sufficient to represent all
states without any truncation.
\begin{verbatim}
State psi1(10,0,FIELD);
State psi2(10,3,FIELD);
State psi3 = 0.5*psi1 - psi2;
psi1 += psi3;
psi3.normalize();
Complex z = psi2*psi3;
\end{verbatim}
The expression {\tt psi1+=psi3} is superior to the alternative {\tt
psi1=psi1+psi3} because it avoids the creation of temporary {\tt State}
objects, which is an important consideration in high-dimensional Hilbert
spaces.

\subsubsection{Operators}   \label{secop}

A general way of representing operators is as $N\times N$ complex matrices
acting on vectors in $N$-dimensional Hilbert space. For large $N$, however,
this can be very inefficient, as these matrices become very large, and applying
them to states requires $O(N^2)$ operations. Fortunately, most of the operators
of interest in quantum systems are {\it sparse}, consisting of sums and
products of a few primary operators. For {\tt FIELD}s, such primary operators
are annihilation and creation operators $\hat a$ and $\hat a^\dagger$ and
position and momentum operators $\hat X$ and $\hat P$; for {\tt SPIN}s, the
primary operators are the Pauli matrices $\hat\sigma_i$; for {\tt ATOM}s, we
have the transition operators $\ket i\langle j|$.

In the program, these primary operators are implemented as simple classes, as
illustrated for the {\tt SPIN} operator $\hat\sigma_+$ in the following code
section.
\begin{verbatim}
class SigmaPlus: public PrimaryOperator {
public:
  SigmaPlus() : PrimaryOperator(0,SPIN) {};
  SigmaPlus(int freedom) : PrimaryOperator(freedom,SPIN) {};
  virtual void applyTo(State&,int,double);
};
void SigmaPlus::applyTo(State& v, int hc, double) {
  switch( hc ) {
  case NO_HC:
    v[1] = v[0]; v[0] = 0; break;
  case HC:
    v[0] = v[1]; v[1] = 0; break; 
  }
}
\end{verbatim}
The {\tt SigmaPlus} class is derived from the abstract class {\tt
PrimaryOperator} which serves as an interface to the different special classes
like {\tt SigmaPlus}. Apart from the two constructors, the class contains only
the method {\tt applyTo}. The three arguments of {\tt applyTo} are a
single-degree of freedom {\tt State}, an integer switch determining whether to
apply $\hat\sigma_+$ or its Hermitian conjugate, and a {\tt double} argument
specifying the time for time-dependent operators, which is not used here.

The program represents {\it composite operators}, i.e., sums and products of
primary operators, by stacks containing
pointers to primary operators as illustrated in Fig.~\ref{figstack}. Those
stacks are the principal member data of the {\tt Operator} class, which is the
parent class of {\tt PrimaryOperator} and therefore of all special classes
derived from {\tt PrimaryOperator}. For a primary operator like {\tt
SigmaPlus}, the stack consists just of the pointer to {\tt *this}, which points
to the primary operator itself. Figure~\ref{figinherit} shows the hierarchy of
operator classes.

The example stack in box 3 in Fig.~\ref{figstack} is generated by the code
segment
\begin{verbatim}
Operator O1 = a + b;
Operator O2 = (3 * c) * d;
Operator O3 = 01 - 02;
\end{verbatim}
where {\tt a, b, c, d} are assumed to be primary operators defined earlier in
the program.  The example illustrates how addition, subtraction, and
multiplication of {\tt Operator}s is implemented in terms of operations on the
stack. Further operations defined for {\tt Operator}s include Hermitian
conjugation and raising to an integer power. The C++ inheritance mechanism
ensures that all these operations are also defined for the derived
primary-operator classes like {\tt SigmaPlus}.

To apply an {\tt Operator} to a {\tt State}, the `{\tt *}' operator can be used
as in the following example, where {\tt psi} is a {\tt State} and {\tt O3} is
defined above:
\begin{verbatim}
State psi1 = O3 * psi;
\end{verbatim}
Internally, this is implemented as a recursive evaluation of the stack.
The order in which the primary operators are applied in the example can be
inferred from the parentheses in
\begin{equation}
\hat O_3\ket\psi = \left(\hat a + \hat b - 3\hat c\hat d\right)\ket\psi
= \left(\hat a\ket\psi + \hat b\ket\psi\right) 
 - \hat c\left(3(\hat d\ket\psi)\right) \;.
\end{equation}
The program keeps the number of operations and the number of temporary
{\tt State}s it creates to a minimum. Some care has to be exercised, however, to
avoid an inefficient evaluation order. E.g., in the code segment
\begin{verbatim}
double x=1.5;
SigmaPlus Sp;
State psi1 = 2.0*x*Sp*psi;
\end{verbatim}
the state {\tt Sp*psi} is first multiplied by 1.5, then by 2.0, whereas in
\begin{verbatim}
double x=1.5;
SigmaPlus Sp;
State psi1 = (2.0*x)*Sp*psi;
\end{verbatim}
there is only one multiplication by 3.0. 

The creation of unnecessary temporary {\tt State}s can be avoided by applying
{\tt Operator}s to {\tt State}s using the `{\tt *=}' operator as in
\begin{verbatim}
SigmaPlus Sp;
State psi(2,SPIN);
psi *= Sp;
\end{verbatim}
When this code segment is executed, no temporary {\tt State}s are created,
in contrast to the otherwise equivalent code segment
\begin{verbatim}
SigmaPlus Sp;  
State psi(2,SPIN);
psi = Sp*psi;
\end{verbatim}

A detailed explanation of the stack and the recursive evaluation procedure
can be found in the extensively commented file {\tt Operator.cc}.

\subsection{Multiple degrees of freedom}

A quantum system with $M$ degrees of freedom can be represented in a product
Hilbert space ${\cal H}_1 \otimes \cdots \otimes {\cal H}_M$. We assume that
there is a finite, perhaps truncated, product basis
$\{\ket{\phi_{n_1}}\otimes\cdots\otimes\ket{\phi_{n_M}}\,|\,1\le n_j\le N_j\}$.
Any state $\ket\psi\in{\cal H}_1 \otimes \cdots \otimes {\cal H}_M$ can then be
written in the form
\begin{equation}
  \ket\psi = \sum_{n_1, \ldots, n_M} C_{n_1, \ldots, n_M}
  \ket{\phi_{n_1}}\otimes\cdots\otimes\ket{\phi_{n_M}} \;,
\end{equation}
requiring a total of $N_{\rm tot} = N_1 N_2 \cdots N_M$ basis vectors.  To
represent a state with multiple degrees of freedom, the {\tt State} class
contains as member data the number of freedoms $M$, an array of $M$ physical
types, an array of $M$ subspace dimensions $N_j$, and an array of $N_{\rm tot}$
amplitudes $C_{n_1,\ldots,n_M}$.

Product states can be initialized by passing a list of single-degree-of-freedom
states to the appropriate {\tt State} constructor. This is illustrated in the
following example, where the state
$\ket0\otimes\ket0\otimes|\!\downarrow\rangle$ is assigned to {\tt psiIni}:
\begin{verbatim}
State phi1(50,FIELD);
State phi2(50,FIELD);
State phi3(2,SPIN);
State stateList[3] = {phi1, phi2, phi3};
State psiIni(3,stateList);
\end{verbatim}
Entangled states can be constructed by adding several product states or by
explicitly initializing the array of amplitudes $C_{n_1,\ldots,n_M}$.

Operators acting on multiple degrees of freedom are represented as sums and
products of primary operators each acting on a single degree of freedom.  Take
the example of a primary operator $\hat b$ acting on the first degree of
freedom. It can be rewritten as the operator $\hat b\otimes\hat1$ on the
product Hilbert space, where $\hat1$ is the identity operator acting
on all the other degrees of freedom. We can write any state $\ket\psi$ as
\begin{equation}
\ket\psi = \sum_{n_2,\ldots,n_M} |\psi_{n_2,\ldots,n_M}\rangle
  \otimes \ket{\phi_{n_2}}\otimes\cdots\otimes\ket{\phi_{n_M}} \;;
\end{equation}
the action of $\hat b\otimes\hat1$ on $\ket\psi$ is therefore given by the
action of $\hat b$ on the first degree of freedom inside a hierarchy of loops
over all the other degrees of freedom:
\begin{equation}
\left(\hat b\otimes{\hat 1}\right) \ket\psi =
  \sum_{n_2,\ldots,n_M} \left(\hat b|\psi_{n_2,\ldots,n_M}\rangle\right)
  \otimes \ket{\phi_{n_2}}\otimes\cdots\otimes\ket{\phi_{n_M}} \;.
\end{equation}
In the program, the loops are unfolded into one big loop if the primary
operator acts on the first or last degree of freedom; otherwise the loops are
unfolded into two loops, an ``inner'' and an ``outer'' loop.

To define, e.g., a primary {\tt SigmaPlus} operator acting on the $n=3$rd
degree of freedom, the constructor has to be called with the argument $n-1=2$:
\begin{verbatim}
SigmaPlus Sp(2);
\end{verbatim}

The {\tt Operator} class is virtually unaffected by the complications arising
  from multiple degrees of freedom (see Fig.~\ref{figmultiple}).
Whenever an {\tt Operator} is applied to a
{\tt State} {\tt psi}, the recursive evaluation of the {\tt Operator} stack
will eventually come across a pointer to some primary operator {\tt B} acting
on a particular freedom. At that stage, the pointer to {\tt B} will be passed
to the method {\tt psi.apply()} of the {\tt State} class, which controls the
loops over all the other degrees of freedom. Each time the loop is executed,
the {\tt State} class passes a single-degree-of-freedom state to the method
{\tt B.applyTo()} of the primary operator {\tt B}. The complex amplitudes of
this single-freedom state are typically stored at widely spaced locations in
the array of complex amplitudes $C_{n_1,\ldots,n_M}$, but this fact is
completely hidden from the primary operator {\tt B}.

This way of organizing the program has great advantages. 
Most importantly, all the
implementation details of multiple-freedom states are hidden from the {\tt
Operator} class. Apart from leading to a transparent program, this makes adding
new primary operators very easy, as was seen in Sec.~\ref{secop}. The
definition of the primary-operator class {\tt SigmaPlus} given there is used
without modification in the multiple-freedom case. 

Our class library realizes its full potential when all operators are sums and
products of a few simple primary operators. Although this situation is
extremely common in many fields, there are important exceptions like the
Coulomb potential. While the program could be adapted to implement such a case,
some of its unique features would be lost in the process.

For efficiency, the {\tt State} class distinguishes internally between
single-freedom and multiple-freedom states; many actions are more efficient for
a single degree of freedom.  This distinction, however, is completely
transparent.  The user need distinguish between the two only when constructing
the initial state.

\subsection{The Moving Basis}

In quantum-trajectory simulations, one often encounters {\tt FIELD} states that
are well {\it localized\/} in phase space
{\cite{Gisin1992c,Diosi1988c,%
Gisin1993b,Percival1994b,Steimle1995a,Holland1996a}}. In cases
with strong localization, it is often possible to reduce drastically the number
$N$ of basis states needed by continually changing the basis. If a state is
localized about a point $(q,p)$ in phase space far from the origin, it requires
many number states $\ket n$ to represent it.  But relatively few 
displaced number states (or {\it excited
coherent states\/}) $|q,p,n\rangle = \hat D(q,p)\ket n$, are needed, with
corresponding savings in computer storage space and computation time.  The
operator $\hat D(q,p)$ is the usual coherent state displacement operator
{\cite{Louisell1973}},
\begin{equation}
\hat D(q,p) = \exp {i\over\hbar} \biggl( p\hat X - q\hat P \biggr) \;,
\end{equation}
where $\hat X$ and $\hat P$ are the position and momentum operators.  The
separation of the representation into a classical part $(q,p)$ and a quantum
part $|q,p,n\rangle$ is called the {\it moving basis\/} {\cite{Schack1995c}}
or, as in \cite{Steimle1995a}, the {\it mixed\/} representation.  To represent
a state of type {\tt FIELD} in the moving basis requires to store the complex
center of coordinates $\alpha=(q+ip)/\sqrt{2}$ in addition to the complex
amplitudes.  A multiple-freedom state in the moving basis with several freedoms
of type {\tt FIELD} requires an array of centers of coordinates.

Implementing the moving basis algorithm is straightforward.  Suppose that at
time $t=t_0$ the state $\ket{\psi(t_0)}$ is represented in the basis
$\ket{q_0,p_0,n}$, centered at
\begin{equation}
(q_0,p_0) = \big(\expect{\psi(t_0)|\hat X|\psi(t_0)},
\expect{\psi(t_0)|\hat P|\psi(t_0)}\big).
\end{equation}
Then after one discrete time step, the expectations in this basis shift to 
\begin{equation}
(q_1,p_1) = \big(\expect{\psi(t_0+\delta t)|\hat X|\psi(t_0+\delta
t)}, 
\expect{\psi(t_0+\delta t)|\hat P|\psi(t_0+\delta t)}\big)\ne
(q_0,p_0).
\end{equation} 
The computational advantage of a small number of basis states is then retained
by changing the representation to the shifted basis $\ket{q_1,p_1,n}$ centered
at $q_1$ and $p_1$.  This shift in the origin of the basis represents the
elementary single step of the moving basis.

The components of $\ket{\psi(t_0+\delta t)}$ can be computed using the
expressions given above. The computing time needed for the basis shift is of
the same order of magnitude as for computing a single discrete time step of one
of the quantum trajectory equations.  Shifting the basis once every discrete
time step could therefore double the computing time, depending on the
complexity of the Hamiltonian and the number of degrees of freedom. On the
other hand, the reduced number of basis vectors needed to represent states in
the moving basis can lead to savings far bigger than a factor of 2.

In the example of second harmonic generation discussed in {\cite{Schack1995c}},
two modes of the electromagnetic field interact. Using the moving basis reduces
the number of basis vectors needed by a factor of 100 in each mode. The total
number of basis vectors needed is thus reduced by a factor of 10000, leading to
reduction in computing time by a factor of roughly $10000/2=5000$. Furthermore,
the fixed basis would exceed the memory capacity of most computers.

The {\tt State} class includes a variety of basis-changing methods. The most
important  is the method
\begin{verbatim}
void moveCoords( const Complex& displacement, int theFreedom,
                 double shiftAccuracy );
\end{verbatim}
which performs a relative shift of the center of coordinates
$\alpha=(q+ip)/\sqrt{2}$ by an amount given by the complex argument {\tt
displacement}. The integer argument {\tt theFreedom} specifies which degree of
freedom is to be shifted---this degree of freedom must be of type {\tt
FIELD}. The {\tt double} argument {\tt shiftAccuracy} gives the numerical
accuracy with which to make the shift. The physical state is unchanged by
applying {\tt moveCoords()}, but it is represented in a new basis. The method
{\tt moveCoords()} is used in the stochastic integration algorithms of the {\tt
Trajectory} class described in Sec.~\ref{sectraj}. The primary operators of
type {\tt FIELD} defined in the files {\tt FieldOp.h} and {\tt FieldOp.cc} are
implemented in such a way that they can handle moving-basis states as well as
ordinary states.

The quantum trajectory equations can contain both localizing and delocalizing
terms. {\cite{Gisin1992c,Diosi1988c,%
Gisin1993b,Percival1994b,Steimle1995a,Holland1996a}}.
Nonlinear terms in the Hamiltonian tend to spread the wave function in
phase space, whereas the Lindblad terms often cause it to
localize. Accordingly, the width of the wave packets varies along a typical
trajectory. We use this to reduce the computing time even further by
dynamically adjusting the number of basis vectors. Our criterion for this
adjustment depends on parameters $\epsilon\ll1$, the {\it cutoff probability},
and $N_{\rm pad}$, the {\it pad size}, which represents the number of boundary
basis states that are checked for significant probability. We require the total
probability of the top $N_{\rm pad}$ states to be no greater than $\epsilon$,
increasing or decreasing the number of states actually used accordingly, as
the integration proceeds along the quantum trajectory.  The method of the {\tt
State} class used to adjust the basis size is
\begin{verbatim}
void adjustCutoff(int theFreedom, double epsilon, int padSize);
\end{verbatim}
where the arguments specify the degree of freedom to be adjusted, the cutoff
probability $\epsilon$, and the pad size $N_{\rm pad}$, respectively.

Like the basis-changing methods discussed above, the method {\tt adjustCutoff()}
is typically only used inside integration routines of the {\tt Trajectory}
class. Those methods will not normally be called from a top-level program,
so the user need not be concerned by them.

\section{The Trajectory class}   \label{sectraj}

The {\tt Trajectory} class and its associated classes, defined in the files
{\tt Traject.h} and {\tt Traject.cc}, implement the integration of the quantum
trajectory equations~(\ref{eqqsd}), (\ref{eqqj}), and (\ref{eqortho}). At the
heart of this part of the code is the abstract class {\tt IntegrationStep}
which serves as an interface for the specific stepper classes implementing
single integration steps of lenght $dt$. The stepper classes derived from the
class {\tt IntegrationStep} include the class {\tt Order4Step} for a single
4-th order Runge-Kutta step of the QSD equation~(\ref{eqqsd}) as well as a
group of classes using adaptive Cash-Karp Runge-Kutta time steps: the class
{\tt AdaptiveStep} for a time step of total length $dt$ of the QSD
equation~(\ref{eqqsd}), the class {\tt AdaptiveJump} for a time step of total
length $dt$ of the quantum jump equation~(\ref{eqqj}), and the class {\tt
AdaptiveOrthoJump} for a time step of total length $dt$ of the orthogonal jump
equation~(\ref{eqortho}). All those classes use a single first order Euler
integration step of length $dt$ for the stochastic part.
Due to the modular structure of the class library, it is straightforward to add 
more sophisticated stochastic integration methods (see, e.g., 
{\cite{Kloeden1992,Steinbach1995a}}). 

To initialize a stepper, including all temporary memory needed for the
integration algorithm, all one has to do is call the appropriate
constructor as in the code segment
\begin{verbatim}
State psiIni(2,SPIN);
SigmaPlus Sp;
Operator H = Sp + Sp.hc();
int nL = 1;
Operator L[nL] = {0.1*Sp.hc()}
AdaptiveStep stepper(psiIni, H, nL, L);
\end{verbatim}
A less trivial example can be found in the sample program in
Sec.~\ref{secsample}.  Entire quantum trajectories are computed by repeatedly
calling a stepper from within the {\tt Trajectory} class. A trajectory is
initialized as in the following example which is taken from the sample program
below:
\begin{verbatim}
double dt=0.01;    // basic time step passed to the stepper
ACG gen(38388389); // random number generator defined in ACG.h
ComplexNormal rndm(&gen); 
                   // Gaussian random numbers defined in CmplxRan.h
Trajectory traj(psiIni, dt, stepper, &rndm);
\end{verbatim}

The {\tt Trajectory} class comprises two methods to launch the
simulation, compute expectation values of operators of interest, and produce
output. The use of the method {\tt plotExp()}, designed to simulate a single
trajectory, is explained in Sec.~\ref{secsample}. The method {\tt sumExp()},
which is very similar to {\tt plotExp()}, can be used to compute the mean
expectation values of operators averaged over many trajectories.

\section{Sample program and template}   \label{secsample}

In this section, we illustrate the main features of the class library in a
complete example program which can be used as a template.  The example program
computes expectation values for a single trajectory of the quantum state
diffusion equation (\ref{eqqsd}); to compute means over many trajectories, one
simply replaces the call to {\tt traj.plotExp()} in the template by a call to 
{\tt traj.sumExp()}.  The system has three degrees of freedom:
two nonlinearly coupled field modes described by annihilation operators $\hat
a_1$ and $\hat a_2$, and a spin described by raising and lowering operators
$\hat\sigma_+$ and $\hat\sigma_-$. The Hamiltonian in the interaction picture
is \cite{Schack1996c}
\begin{equation}
\hat H = Ei(\hat a_1\d-\hat a_1) + {\chi\over2}i(\hat a_1^{\dagger2}\hat a_2 
    - \hat a_1^2\hat a_2\d) + \omega\hat\sigma_+\hat\sigma_-
    + \eta i(\hat a_2\hat\sigma_+  - \hat a_2\d\hat\sigma_-) \;,
\end{equation}
where $E$ is the strength of an external pump field, $\chi$ is the strength of
the nonlinear interaction, $\omega$ is the detuning between the frequency of
the field mode $\hat a_2$ and the spin transition frequency, and $\eta$ is
the strength of the coupling of the spin to the field mode $\hat a_2$. The
Lindblad operators
\begin{equation}
\hat L_1=\sqrt{2\gamma_1}\,\hat a_1 \;,\;\;\;
\hat L_2=\sqrt{2\gamma_2}\,\hat a_2 \;,\;\;\;
\hat L_3=\sqrt{2\kappa}\,\hat\sigma_-
\end{equation}
describe dissipation of the field modes and the spin with coefficients
$\gamma_1$, $\gamma_2$, and $\kappa$, respectively. 

The trajectory's initial state is the product state $\ket{\psi_{\rm
ini}}=\ket0\otimes\ket0\otimes|\!\downarrow\rangle$. The integration step-size
is {\tt dt=0.01} and the total integration time is {\tt 500*dt = 5}.  The
integration stepper {\tt AdaptiveStep} implements a single time step of length
{\tt dt} of the QSD equation~(\ref{eqqsd}) using the Cash-Karp Runge-Kutta
algorithm with adaptive time steps {\cite{Press1992}} for the deterministic
part and first-order Euler integration for the stochastic part.

At times that are integer multiples of {\tt 50*dt = 0.5}, the expectation
values of the operators specified in the array {\tt outlist} are computed and
written to the files specified in the array {\tt flist}. E.g., the first
element of {\tt outlist} is the operator $\hat X_1\equiv\hat\sigma_+\hat
a_2\hat\sigma_-\hat\sigma_+$. At times $t=0,0.5,\ldots,5.0$, the method {\tt
plotExp} computes the expectation values $\expect{\hat X_1}$ and ${\rm
var}(\hat X_1)\equiv\expect{\hat X_1-\expect{\hat X_1}}$ and writes $t$, ${\rm
Re}(\expect{\hat X_1})$, ${\rm Im}(\expect{\hat X_1})$, ${\rm Re}({\rm
var}(\hat X_1))$, and ${\rm Im}({\rm var}(\hat X_1))$ to the file {\tt
X1.out}. In addition, each time a set of expectation values is computed, the
program writes 7 numbers to standard output (see the sample output below): the
time $t$, 4 expectation values determined by the integer array {\tt pipe}, the
number of basis states used, and the number of adaptive steps taken. The
integers in the array {\tt pipe} correspond to the columns in the output files
containing expectation values (i.e., columns 2 through 5 of each output file).
In the present example, expectation values are computed for the 5
operators $\hat\sigma_+\hat a_2\hat\sigma_-\hat\sigma_+$, 
$\hat\sigma_-\hat\sigma_+\hat a_2\hat\sigma_-$, $\hat a_2$, $\hat n_1$, 
and $\hat n_2$, which are written to 5 output files with numbered columns 1
through 20. According to the expression {\tt int pipe[]=\{1,5,13,17\}}, the
expectation values written to standard output are 
${\rm Re}(\expect{\hat\sigma_+\hat a_2\hat\sigma_-\hat\sigma_+})$, 
${\rm Re}(\expect{\hat\sigma_-\hat\sigma_+\hat a_2\hat\sigma_-})$, 
${\rm Re}(\expect{\hat n_1})$, and 
${\rm Re}(\expect{\hat n_2})$.

The moving basis is used for both {\tt FIELD} degrees of freedom. The basis
size is dynamically adjusted with a cutoff probability $\epsilon=0.01$ and a pad
size $N_{\rm pad}=2$. The sample output below shows how the basis size changes
with time. Initially, $5000=50*50*2$ states are allocated, but at time $t=0.5$,
only 18 states are needed. Subsequently, the basis size fluctuates around a
typical size of 70 states.

Here is the complete program:

\begin{verbatim}
#include "Complex.h"
#include "ACG.h"
#include "CmplxRan.h"
#include "State.h"
#include "Operator.h"
#include "FieldOp.h"
#include "SpinOp.h"
#include "Traject.h"

int main() 
{
// Primary Operators
  AnnihilationOperator A1(0);  // 1st freedom
  NumberOperator N1(0);
  AnnihilationOperator A2(1);  // 2nd freedom
  NumberOperator N2(1);
  SigmaPlus Sp(2);             // 3rd freedom
  Operator Sm = Sp.hc();       // Hermitian conjugate
  Operator Ac1 = A1.hc();
  Operator Ac2 = A2.hc();
// Hamiltonian
  double E = 20.0;           
  double chi = 0.4;      
  double omega = -0.7;       
  double eta = 0.001;
  Complex I(0.0,1.0);
  Operator H = (E*I)*(Ac1-A1)
             + (0.5*chi*I)*(Ac1*Ac1*A2 - A1*A1*Ac2)
             + omega*Sp*Sm + (eta*I)*(A2*Sp-Ac2*Sm);
// Lindblad operators
  double gamma1 = 1.0;       
  double gamma2 = 1.0;       
  double kappa = 0.1;        
  const int nL = 3;
  Operator L[nL]={sqrt(2*gamma1)*A1,sqrt(2*gamma2)*A2,sqrt(2*kappa)*Sm};
// Initial state
  State phi1(50,FIELD);       // see Section 4.2
  State phi2(50,FIELD);
  State phi3(2,SPIN);
  State stateList[3] = {phi1,phi2,phi3};
  State psiIni(3,stateList);
// Trajectory
  double dt = 0.01;    // basic time step                            
  int numdts = 50;     // time interval between outputs = numdts*dt  
  int numsteps = 10;   // total integration time = numsteps*numdts*dt
  int nOfMovingFreedoms = 2;
  double epsilon = 0.01;     // cutoff probability
  int nPad = 2;              // pad size
  ACG gen(38388389);         // random number generator with seed
  ComplexNormal rndm(&gen);  // Complex Gaussian random numbers
  AdaptiveStep stepper(psiIni, H, nL, L);       // see Section 5
  Trajectory traj(psiIni, dt, stepper, &rndm);  // see Section 5
// Output
  const int nOfOut = 5;
  Operator outlist[nOfOut]={ Sp*A2*Sm*Sp, Sm*Sp*A2*Sm, A2, N1, N2 };
  char *flist[nOfOut]={"X1.out","X2.out","A2.out","N1.out","N2.out"};
  int pipe[] = { 1, 5, 13, 17 };    // controls standard output
// Simulate one trajectory
  traj.plotExp( nOfOut, outlist, flist, pipe, numdts, numsteps,
                nOfMovingFreedoms, epsilon, nPad );
}
\end{verbatim}
In addition to the output files {\tt X1.out}, {\tt X2.out}, {\tt
A2.out}, {\tt N1.out}, and {\tt N2.out}, the program writes the
following lines to standard output:
\begin{verbatim}
0    0            0            0        0      5000    0
0.5  0.000505736  0.000504849 52.3875   3.5807   18   75
1    0.0131402    0.0131173   51.8747  35.1089   60   50
1.5  0.0329714    0.0320222   32.8707  44.3184  108   50
2    0.0425276    0.0455457   32.1562  41.7798   70   56
2.5  0.0284912    0.0564117   34.85    37.8809   80  117
3    0.0260639    0.0626976   33.9828  39.3437   80  143
3.5  0.0544306    0.0439029   51.0632  37.6462   70   99
4    0.0796275   -0.0209383   41.9614  38.0884   70  167
4.5  0.0834672   -0.0543796   33.1194  36.1007   70  195
5   -0.00616844   0.0110794   76.6321  29.4303   50  119
\end{verbatim}

\section*{Acknowledgements}

We would like to thank I. C. Percival for initiating the project, for 
pushing us in the right direction, and for invaluable discussions. We
would also like to thank M. Rigo for several improvements of the code,
and most notably for implementing the quantum jump and orthgonal jump
classes. Financial support was provided by the UK EPSRC. 

%\bibliographystyle{prsty}
%\bibliography{/home/rschack/lit/p}

\newpage

\begin{figure}   % ~brun/Talks/Code/Paper/stack.ps
\caption{These examples show how the internal stack representations of primary
and composite {\tt Operator}s are combined in arithmetic operations.  Notice
that while arithmetic expressions are parsed from left to right, the order in
which {\tt Operator}s are applied to {\tt State}s is from right to left.}
\label{figstack}
\end{figure}

\begin{figure}   % ~brun/Talks/Code/Paper/operator.ps
\caption{In this diagram, the arrows point from parent classes to derived
classes. The classes listed in each box are declared in the {\tt \#include} file
given above the box. Arithmetic operations are defined in the {\tt
Operator} class. The {\tt PrimaryOperator} class serves as an interface for the
specific {\tt FIELD}, {\tt SPIN}, and {\tt ATOM} operators. Adding operators of
either an existing or a new type is straightforward.}
\label{figinherit}
\end{figure}

\begin{figure}   % ~brun/Talks/Code/Paper/state2.ps
\caption{When a pointer to a primary operator acting on a particular freedom is
encountered during the evaluation of an {\tt Operator} stack, control is passed
to the {\tt State} class, where within loops over the basis
states of all the other degrees of freedom, the primary operator is applied
to a succession of single-freedom
states. This means that the {\tt Operator} class and its derived classes do not
need to distinguish between single and multiple freedom states; all details
concerning the multiple-freedom case are hidden within the {\tt State} class.}
\label{figmultiple}
\end{figure}

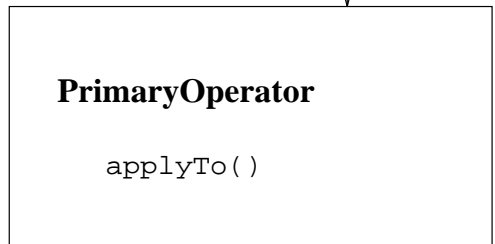
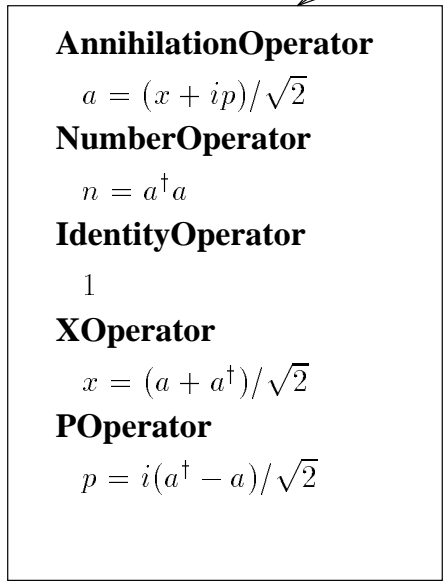
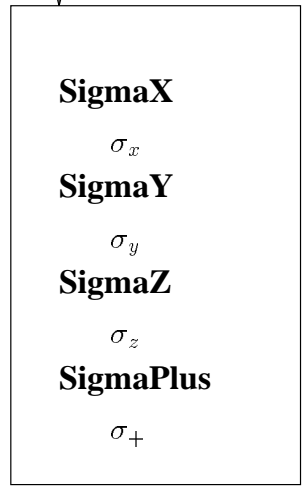
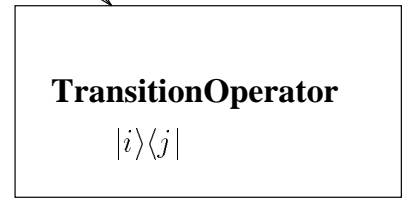
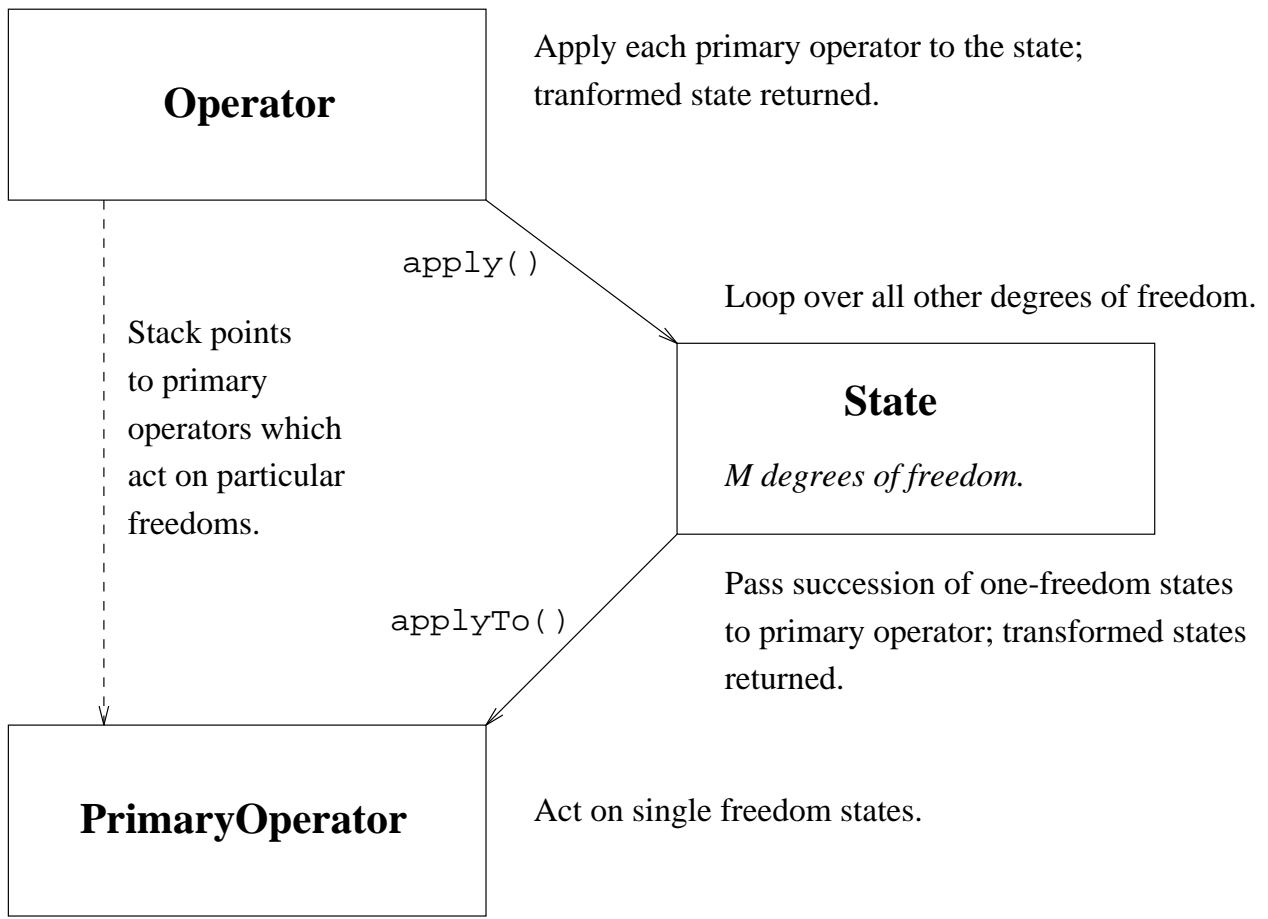
\end{document}